\documentclass[10pt]{article}
\usepackage{cite}
\usepackage{epsfig}
\usepackage[dvips,usenames]{color}
\usepackage{color}
\usepackage{rotating}
\usepackage{amsmath,amssymb,amsfonts}
\usepackage{latexsym}
\begin{document}
\setcounter{section}{0}
\setcounter{equation}{0}
\setcounter{figure}{0}
\setcounter{table}{0}
\setcounter{footnote}{0}
\begin{center}
{\bf\Large Classes of $f$-Deformed Landau Operators:}

\vspace{5pt}

{\bf\Large Nonlinear Noncommutative Coordinates}

\vspace{5pt}

{\bf\Large from Algebraic Representations}\footnote{Contribution to the Proceedings of the Fifth International
Workshop on Contemporary Problems in Mathematical Physics, Cotonou, Republic of Benin, October 27--November 2, 2007,
eds. Jan Govaerts and M. Norbert Hounkonnou (International Chair in Mathematical Physics and Applications,
ICMPA-UNESCO, Cotonou, Republic of Benin, 2008), pp.~124--129.}
\end{center}
\vspace{10pt}
\begin{center}
Joseph BEN GELOUN$^{\dagger}$, Jan GOVAERTS$^{\ddagger,\star,\dagger}$
%\footnote{Fellow of the Stellenbosch Institute for Advanced Study (STIAS), Stellenbosch, South Africa.}
and M. Norbert HOUNKONNOU$^{\dagger}$\\
\vspace{5pt}
$^\dagger${\sl International Chair in Mathematical Physics and Applications (ICMPA-UNESCO Chair),\\
University of Abomey--Calavi, 072 B. P. 50, Cotonou, Republic of Benin}\\
{\it E-Mail: joseph.bengeloun@cipma.uac.bj, norbert.hounkonnou@cipma.uac.bj}\\
\vspace{7pt}
$^\ddagger${\sl Center for Particle Physics and Phenomenology (CP3),\\
Institut de Physique Nucl\'eaire, Universit\'e catholique de Louvain (U.C.L.),\\
2, Chemin du Cyclotron, B-1348 Louvain-la-Neuve, Belgium}\\
{\it E-Mail: Jan.Govaerts@uclouvain.be}\\
\vspace{7pt}
$^\star${\sl Fellow, Stellenbosch Institute for Advanced Study (STIAS),\\
7600 Stellenbosch, Republic of South Africa}
\end{center}

\vspace{15pt}

\begin{quote}
We consider, in a superspace, new operator dependent
noncommutative (NC) geometries of the nonlinear quantum Hall
limit related to classes of $f$-deformed Landau operators
in the spherical harmonic well. Different NC coordinate algebras
are determined using unitary representation spaces of
Fock-Heisenberg tensored algebras and of the Schwinger-Fock
realisation of the $su(1,1)$ Lie algebra.
A reduced model allowing an underlying ${\mathcal N}=2$ superalgebra
is also discussed.
\end{quote}

\vspace{10pt}

\section{Introduction}
\label{JBN.Sect1}

Generalised Landau problems have recently been studied
in the context of deformed quantum mechanics\cite{BGH1,BGH5}.
It has been shown that, beyond the well-known duality between
interactions and noncommutative (NC) geometry\cite{scholtz},
the deformed quantum Hall NC geometry becomes fully operator
dependent.

In this contribution, we pursue our previous investigations\cite{BGH1}
of generalised NC coordinate algebras stemming from effects
of interactions and algebraic extensions of canonical quantum mechanics.
Extending the deformed Landau model\cite{BGH1} to a superspace,
we consider a $f$-deformed quantisation, \`a la
Jannussis {\it et al.\/}\cite{jannus}, of a Landau operator
in a spherical well with an additional odd Grassmann
harmonic oscillator. 
It ought to be emphasized that the choice of the bosonic
and fermionic harmonic potentials fits the exact solvability
condition of the problem in the presence of an electromagnetic field.
Furthermore, the NC harmonic strength has to
be related to a nontrivial time reversal symmetry
breaking\cite{schol} when the given system is only
composed of constant piecewise potentials and of configurations
with negative angular momentum values for some states.

The outline is as follows.
Section 2, in which notations are specified, is devoted to
the study of the classical model.
In Section 3, the $f$-deformed theory is considered. 
Projecting the coordinates onto infinite
dimensional sub-Hilbert spaces playing the r\^ole of
specific types of algebra representation spaces,
we derive NC geometries associated with the modified
model. A reduced class of models allows us to deduce
the existence of a ${\mathcal N}=2$ superalgebra.
The paper ends with some remarks in Section 4.

\section{Classical Study}
\label{JBN.Sect2}

Consider the nonrelativistic confined motion of a particle
of mass $m$ and charge $q$ with
$\vec{r}=(x,y,0)$ as position coordinate vector
in some planar nanoscopic semiconductor sample subjected to static background fields,
 namely, a planar electric field $\vec{E}$
and a magnetic field $\vec{B}$ perpendicular to the $(x,y)$-plane.
In addition, the particle is submitted to a spherical
harmonic potential with angular frequency $\omega$
and stiffness constant $k=m\omega^{2}$.
Finally, we introduce an odd Grassmann harmonic oscillator
with angular frequency $\omega_{0}$ and odd Grassmann degrees of freedom
$\theta$ and $\theta^{\dag}$.
Units such that $c=1$ are used implicitly.

The total Lagrange function $L$ of the system consists of the sum of
two functions, $L_0$ and $L_1$, given as follows in the symmetric gauge
$\vec{A}(\vec{r}\,)=\vec{B}\times\vec{r}/2$,
\begin{eqnarray}
\label{lag}
L&=& L_0 + L_1,  \nonumber \\
 && \nonumber \\
L_0&=&\textstyle {\frac{1}{2}}\,m\,\dot{\vec{r}}^{\;2}
+ q\,(\vec{E}\cdot \vec{r} + \dot{\vec{r}}\cdot \vec{A}(\vec{r}\,))
 - {\frac{1}{2}}\,m\,\omega^{\;2} \vec{r}^{\;2}, \nonumber \\
 && \nonumber \\
L_1&=&\textstyle  {\frac{i}{2}}\,m_{0}\,\omega_{0}
(\theta^{\dag}\,\dot{\theta} - \dot{\theta}^{\dag}\theta)
- {\frac{1}{2}}\,m_{0}\,\omega^2_{0}\theta^{\dag}\theta.
\end{eqnarray}
In terms of the shifted coordinates defined by
$\vec{R}(t)= \vec{r}(t)- q\vec{E}/k$
which minimise the total potential energy in the radial sector,
we have the momenta conjugate to
$\vec{R}$, $\theta$ and $\theta^{\dag}$, respectively (using left derivatives
with respect to odd Grassmann variables),
\begin{displaymath}
\vec{P}= m\dot{\vec{R}} - \frac{1}{2}q(\frac{q}{k}\,\vec{E} + \vec{R})\times\vec{B},\qquad
\pi_{\theta}= -\frac{i}{2}m_{0}\omega_{0}\,\theta^{\dag},\qquad
\pi_{\theta^{\dag}} =  -\frac{i}{2}m_{0}\omega_{0}\,\theta.
\end{displaymath}
Noting that $L_1$ is already in Hamiltonian form, the canonical Hamiltonian of the system is given as,
\begin{equation}
\label{JBG.ham}
H=\textstyle {\frac{1}{2m}}\left[\vec{\pi}
 - q \vec{A}(\vec{R})\right]^{\;2}
 + {\frac{1}{2}}k\, \vec{R}^{\;2} - {\frac{q ^2}{2k}}\vec{E}^{\;2}
 + {\frac{1}{2}}k_{0}\,\theta^{\dag}\theta,
\end{equation}
where
\begin{equation}
\vec{\pi}:= \textstyle \vec{P} + {\frac{q^{2}}{2k}}\,\vec{E}\times\vec{B},
\qquad \vec{A}(\vec{R})={\frac{1}{2}}\vec{B}\times\vec{R},
\qquad k_{0}:= m_{0}\omega_{0}^{2},
\end{equation}
having introduced the parametrisation $\vec{R}=(X,Y,0)$, $\vec{P}=(P_X,P_Y,0)$ and
$\vec{\pi}=(\pi_{X},\pi_{Y},0)$.
The reduced Poisson brackets for even $(e)$ and
odd $(o)$ quantities are given by,
\begin{displaymath}
\{ Z,P_Z\}_{e}=1=\{ Z,\pi_Z\}_{e},\quad Z=X,Y,\qquad
\{\theta, \theta^{\dag}\}_{o}=-i/(m_{0}\omega_{0}).
\end{displaymath}
By changing variables to classical Fock modes,
we obtain, with $\Omega^{2} = \sqrt{\omega^{2} + \varpi^{2}}$
and $ \varpi = (qB)/(2m)$,
\begin{eqnarray}
&a_{Z} = \left(\frac{m\Omega}{2\hbar}\right)^{1/2}
  [Z + {\frac{i}{m\Omega}}\pi_{Z}],\qquad
a_{\pm} = \frac{1}{\sqrt{2}}[a_{X} \mp i a_{Y}],\qquad
\{a_{\pm}, a_{\pm}^{\ast}\} = -\frac{i}{\hbar},  \\ \cr
& b = \left(\frac{m_{0} \omega_{0}}{2\hbar}\right)^{1/2}\theta,
\qquad
 b^{*} = \left(\frac{m_{0} \omega_{0}}{2\hbar}\right)^{1/2}\theta^{\dag},\quad
\{b,b^{*}\} = -\frac{i}{\hbar},
\label{eq:bbdag}
\end{eqnarray}
$a^{\ast}_\pm$ and $b^{*}$ being the complex conjugates
of the variables $a_\pm$ and $b$, respectively.
After some calculations from (\ref{JBG.ham}),
one finds the reparametrised Hamiltonian,
\begin{eqnarray}
\label{hamo}
H&=&\ \ \textstyle {\frac{1}{2}\hbar\Omega}
(a_{+}a_{+}^{\ast} + a_{+}^{\ast}a_{+}
\,+\, a_{-}a_{-}^{\ast}+ a_{-}^{\ast}a_{-})\cr
&& \cr
&&-\textstyle  {\frac{1}{2}\hbar \varpi}
(a_{+}a_{+}^{\ast} + a_{+}^{\ast}a_{+}
\,-\, (a_{-}a_{-}^{\ast} + a_{-}^{\ast}a_{-}))
\,+\,{\frac{1}{2}\hbar\omega_{0}}\, b^{*}b
\,-\, {\frac{(q\vec{E})^{\;2}}{2k}},
\end{eqnarray}
where $\hbar$ is regarded simply at this stage as a numerical constant introduced
for dimensional convenience.

\section{$f$-Deformations of Landau Operators}
\label{JBN.Sect3}

In this Section, we consider  $f$-deformed classes
of the Landau Hamiltonian function (\ref{hamo}).

\subsection{Deformed models and spectra}
\label{JBN.Subsect31}

Generalising the canonical helicity mode quantisation
given by two usual commuting Fock algebras,
$[a_{\pm},a_{\pm}^{\, \dag}]=1$,
with number operators $N_{\pm}=a^{\, \dag}_{\pm}a_{\pm}$,
we introduce two generalised deformed Fock algebras of
the type constructed by
Jannussis {\it et al.\/}\cite{jannus}.
These algebras are factorised with respect to helicity
and are defined by two real functions $f_{\pm}(N_{\pm})\neq 0$
and the generators $A_{\pm}$, $A_{\pm}^{\, \dag}$ such that,
%\clearpage
\begin{eqnarray}
\label{genalg}
&&A_{\pm}=a_{\pm} f_{\pm}(N_{\pm}),
\qquad A_{\pm}^{\, \dag}=f_{\pm}(N_{\pm})a_{\pm}^{\, \dag},
\cr
&& \cr
&&
[A_{\pm}, A_{\pm}^{\, \dag}]= (N_{\pm}+1)f_{\pm}^{2}(N_{\pm}+1)
-(N_{\pm})f_{\pm}^{2}(N_{\pm}).
\end{eqnarray}
Each of these $f_{\pm}(N_{\pm})$-dependent algebras generalises the
usual Fock algebra obtained for $f_{\pm}(N_{\pm})=\mathbb{I}$.
For notational convenience, we introduce
\begin{displaymath}
\{N\}_{\pm}:=N_{\pm}f^{2}_{\pm}(N_{\pm})=A_{\pm}^{\, \dag}A_{\pm},\qquad
\{N+1\}_{\pm}:=(N_{\pm}+1)f^{2}_{\pm}(N_{\pm}+1)=A_{\pm}A_{\pm}^{\,\dag}.
\end{displaymath}
For the odd Grassmann sector,
we use the ordinary anticommuting algebra $\{b,b^{\dag}\}=\mathbb{I}$,
defining fermionic operators $b$ and $b^{\dag}$.

These algebras and operators have well defined
representations on the Fock Hilbert space ${\mathcal F}$
of states defined by the tensor product of two copies, $F_{b,\pm}$,
associated with the bosonic sector, and the two dimensional representation
space, $F_{f}$, of the fermionic oscillator algebra provided by the spin
orthonormalised states $|0\rangle$ and $|1\rangle$. We have,
\begin{displaymath}
{\mathcal F}=F_{b,+}\otimes F_{b,-}\otimes F_{f}
=\{|n_+,n_-,s\rangle, \;\; n_\pm \in \mathbb{N},\, s=0,1\}.
\end{displaymath}

We restrict the study to the following $f_{\pm}$-deformed quantum Hamiltonians,
\begin{eqnarray}
{\mathcal H}&=&\ \  \textstyle\frac{1}{2}\hbar\Omega
\sum_{\epsilon=\pm}\{A_{\epsilon},A^{\, \dag}_{\epsilon}\}
- {\frac{1}{2}}\hbar\varpi
\sum_{\epsilon=\pm}
\epsilon \{A_{\epsilon},A^{\, \dag}_{\epsilon}\} \cr
 && \cr
&&+ \textstyle\frac{1}{4}\hbar\omega_{0}
\sum_{\epsilon=\pm}\, [A_{\epsilon},A^{\, \dag}_{\epsilon}]\,
 b^{\, \dag}b
-{\frac{q^2}{2k}}\vec{E}^{\;2} + K_0 ,
\label{eq:hpqlan}
\end{eqnarray}
\begin{eqnarray}
{\mathcal H}_{0}&=&\ \ 
\textstyle \hbar\Omega
 \left[ A^{\, \dag}_{+} A_{+} +
A^{\, \dag}_{-} A_{-} \,+\, 1\right]
\,-\, \hbar\varpi \left[ A^{\, \dag}_{+} A_{+} -
A^{\, \dag}_{-} A_{-} \right] \cr
 && \cr
 &&+  \textstyle
{\frac{1}{2}}\hbar\omega_{0}\, [A_{-},A^{\, \dag}_{-}]\,
 b^{\, \dag}b
\,-\, {\frac{q^2}{2k}}\vec{E}^{\;2} + K_0.
\label{eq:hpqlan2}
\end{eqnarray}
The Hamiltonian ${\mathcal H}$ has manifest rotational symmetry,
while a reduced model which generates a superalgebra
may be defined from ${\mathcal H}_{0}$.
In the following, we focus our interest on
${\mathcal H}$ and deal with ${\mathcal H}_{0}$ in Subsection~\ref{JBN.Subsect33}.

The Hamiltonian ${\mathcal H}$ is diagonal in the basis
$\{\,|n_+,n_-,s\rangle\,\}$, $s=0,1$, with eigenvalues,
\begin{eqnarray}
E(n_{+},n_{-},0)&=&\ \ \textstyle  \frac{1}{2}\hbar\Omega
\sum_{\epsilon=\pm}\,
\left( \{n + 1\}_{\epsilon}\, + \, \{n\}_{\epsilon}\right)
\cr
&& \cr
&&- \textstyle  \frac{1}{2}\hbar\varpi
\sum_{\epsilon=\pm}\,
\epsilon\left( \{n + 1\}_{\epsilon}\, + \, \{n\}_{\epsilon}\right)
\,-\, {\frac{q^2}{2k}}\vec{E}^{\;2}+K_0,
\end{eqnarray}
\begin{eqnarray}
E(n_{+},n_{-},1)&=&\ \ 
\textstyle {\frac{1}{2}}\hbar\Omega
\sum_{\epsilon=\pm}\,
\left( \{n + 1\}_{\epsilon}\, + \, \{n\}_{\epsilon}\right)
\cr
&& \cr
&&-\textstyle  \frac{1}{2}\hbar\varpi
\sum_{\epsilon=\pm}\,
\epsilon\left( \{n + 1\}_{\epsilon}\,+ \, \{n\}_{\epsilon}\right)
\cr
&& \cr
&&+\textstyle \frac{1}{4}\hbar\omega_{0}
\sum_{\epsilon=\pm}\,\left( \{n + 1\}_{\epsilon}\, -
\, \{n\}_{\epsilon}\right)
\,-\, {\frac{q^2}{2k}}\vec{E}^{\;2}+K_0,
\label{eq:spectrum}
\end{eqnarray}
with $\{n\}_\epsilon= n_\epsilon f_\epsilon^2(n_\epsilon)$.
The degeneracy can easily be determined
for particular relevant reductions\cite{BGH5}. This is the case
for the model reduced under $k=0$ and $\vec{E}=\vec{0}$,
leading to $\varpi=\Omega$, $\omega_{0}=0$, with the spectrum
$E(n_{+},n_{-},s)$ $=\, \hbar\varpi
\left( \{n + 1\}_{-}\, + \, \{n\}_{-}\right)$, $s=0,1$.
Projecting onto one of the spin subsectors,
the deformed Landau problem remains infinitely degenerate
in the angular momentum $\ell=n_+-n_-\ge -n_-$
for each of the Landau levels distinguished by $n_-\ge 0$.
It is instructive\cite{BGH1} to consider the angular momentum operator
as function of the ordinary numbers $N_\pm$ and not
as the nonlinear combinations
$a^\dag_{\pm} a_{\pm} + a_{\pm} a^{\, \dag}_\pm
=\{N\}_{\pm} + \{N+1\}_{\pm}$,
although this is actually a matter of choice.
Thus the deformed angular momentum operator could
be defined by  $L=\hbar\left(N_+-N_-\right)$,
thence ${\mathcal H}$ and $L$ form a complete set
of commuting operators simultaneously diagonalisable.

\subsection{Nonlinear noncommutative coordinates from algebraic representations}
\label{JBN.Subsect32}

\subsubsection{Projection onto deformed Landau levels}
\label{JBN.Subsubsect321}

Due to the presence of spin degrees of freedom,
a number of projectors onto the Landau level $n^0_-$
are available. The following projectors prove to be relevant,
\begin{eqnarray}
&&\textstyle   {\mathcal P}(s,n^0_-)=
\sum_{n_+=0}^{\infty}
|n_+,n^0_-, s\rangle \langle n_+,n^0_-, s|,
\qquad s=0,1,
\label{eq:minpro}\\
 && \cr
&&{\mathcal P}(n^0_-)
={\mathcal P}(0, n^0_-) +  {\mathcal P}(1, n^0_-),
\qquad \mathbb{P}(n^0_-)=
 {\mathcal P}(0, n^0_-) +  {\mathcal P}(1, n^0_- - 1).
 \;\;
\label{eq:projectors}
\end{eqnarray}
For any operator $T$ and $s=0,1$, let us adopt the notations,
\begin{equation}
\bar{T}^{\,s}
={\mathcal P}(s, n^0_-)\,\,T\,\,{\mathcal P}(s, n^0_-),
\qquad s=0,1;\qquad
\bar{T}={\mathcal P}(n^0_-)\,\,T\,\,{\mathcal P}(n^0_-),\qquad
\bar{\bf T}=\mathbb{P}(n^0_-)\,\,T\,\,\mathbb{P}(n^0_-).
\label{eq:not}
\end{equation}
On the other hand, the deformed realisations of
the quantum cartesian plane coordinates can be cast into the form,
\begin{eqnarray}
&&\textstyle {\mathcal X}=\frac{1}{2}
\left(\frac{\hbar}{m\Omega}\right)^{1/2}
(A_{+}+A_{-}+A_{+}^{\, \dag}+A_{-}^{\, \dag}),\\
 && \cr
&&\textstyle {\mathcal Y} =\frac{i}{2}
\left(\frac{\hbar}{m\Omega}\right)^{1/2}
(A_{+}-A_{-}-A_{+}^{\, \dag}+A_{-}^{\, \dag}).
\end{eqnarray}
Before projecting onto any given Landau
level, the quantum cartesian plane coordinates ${\mathcal X}$
and ${\mathcal Y}$ do not commute as operators
as expected from general deformed theories\cite{BGH1,BGH5}.
Indeed, from a direct evaluation the commutator
\begin{eqnarray}
\label{eq:xy}
\textstyle
[{\mathcal X},{\mathcal Y}]=\frac{i}{2}\frac{\hbar}{m\Omega}
\left( \{N\}_{+}\,-\, \{N+1\}_{+} + \{N+1\}_{-}-\{N\}_{-}\right)
\end{eqnarray}
is seen not to vanish, unless one effectively sets $f_\pm(N_{\pm})=1$.

Quantum geometry associated with the deformed quantum Hall effect
can be derived from the commutator of the cartesian
coordinate operators projected onto any of the Landau levels.
After some calculations, we obtain, for $s=0,1$,
\begin{equation}
\textstyle
\left[\bar{{\mathcal X}}^{\,s},\bar{{\mathcal Y}}^{\,s}\right]
=\frac{i\hbar}{2m\Omega}
\left(\{N\}_{+}- \{N+1\}_+\right)\,{\mathcal P}(s,n^0_-)
\label{pxp}
\end{equation}
which radically differs from the unit operator
and also from the general quantum plane noncommutativity
(\ref{eq:xy}).
In the situation such that $f_\pm(N_\pm)=1$,
indeed the result characteristic of the quantum Hall effect
in the plane is recovered, namely
$\left[\bar{{\mathcal X}}^{\,s},\bar{{\mathcal Y}}^{\,s}\right]
= -(i\hbar)/(2m\Omega)\, {\mathcal P}(s,n^0_-)$.
If an alternative is considered through the use
of the total spin state projectors
${\mathcal P}(n^0_-)$  and $\mathbb{P}(n^0_-)$,
one easily obtains the corresponding commutators.
In the specific instance such that $\vec{E}=\vec{0}$, $k=0$ and
$\omega_{0}=0$, one then has $\left[x,y\right]=-(i\hbar)/(q B)$.

Actually, projection onto the lowest Landau level
is not obligatory\cite{scholtz,BGH1}
in order to define a relevant NC geometry.
The pertinent requirement is
the separation of energy levels. 
In the following, we will
consider the $su(1,1)$ algebraic representation spaces
from which different NC geometries of the same model
could be revealed.

\subsubsection{Projection onto $su(1,1)$ representation spaces}
\label{JBN.Subsubsect322}

The Schwinger realisation of the $su(1,1)$ algebra is defined
by the two mode operators
$J_{+}= a_{+}^{\dag}a_{-}^{\dag}$, $J_{-}= a_{+}a_{-}$
and $J_{0} = (1/2)\left[N_++ N_- +1 \right]$,
which satisfy the relations
\begin{equation}
[J_{0}, J_{\pm}] =\pm  J_{\pm},\qquad [ J_{+}, J_{-}] = -2 J_{0}.
 \end{equation}
The Casimir operator takes the form $C_2 = -J_-J_+ + J_0(J_0+1)$
and the total number operator can be written
as $L= N_+ - N_-$. These two last operators commute
and can be diagonalised on a common vector basis.
Denoting the eigenvalues of $C_2$ by $C_2=\ell(\ell+1)$,
then $L=-(2\ell + 1)$ since one may show that
$C_2= (1/4)(L+1)(L-1)$. As a consequence, the
$su(1,1)$ unitary representation spaces
are characterised by half integers $\ell \in \mathbb{Z}/2$.
We get the following spaces, given $k_\ell^0=\max(-2\ell,1)$,
\begin{equation}
{\mathcal E}_\ell= \left\{\, |k-1,\,2\ell+k\rangle, \;
 k=k_\ell^0,k_\ell^0 +1,k_\ell^0 +2,\ldots \,\right\},
\end{equation}
and the relations
\begin{eqnarray}
&& J_0|k-1,\,2\ell+k\rangle = (M= \ell+k)\,|k-1,\,2\ell+k\rangle,\\
&& \cr
&&  C_2|k-1,\,2\ell+k\rangle = \ell(\ell+1)\,|k-1,\,2\ell+k\rangle,\\
&& \cr
&& J_-|k-1,\,2\ell+k\rangle =\sqrt{(k-1)(2\ell+k)}\,
|k-2,\,2\ell+k-1\rangle,\\
&& \cr
&& J_+|k-1,\,2\ell+k\rangle =\sqrt{k(2\ell+k+1)}\,
|k,\,2\ell+k+1\rangle.
\end{eqnarray}
For $s\in \{0,1\}$ and $\ell\in \mathbb{Z}/2$,
a family $\{ P(s,\ell)\}_{s,\ell}$ of projectors
onto the Hilbert subspaces
${\mathcal E}_\ell\otimes|s\rangle \subset {\mathcal F}$
is given by,
\begin{equation}
 \textstyle P(s,\ell)=
\sum_{k=k_\ell^0}^{\infty}
|k-1,2\ell+k, s\rangle \langle k-1,2\ell+k, s|.
\label{eq:psl}
\end{equation}
Let $\epsilon=\pm 1$ be a new parameter.
We can form the linear combination of the spaces
${\mathcal E}_\ell$ and ${\mathcal E}_{\ell+ \frac{\epsilon}{2} }$.
Now, by further combining the operators (\ref{eq:psl}), we get
a new set of projectors onto the space
$({\mathcal E}_\ell \oplus {\mathcal E}_{\ell+ \frac{\epsilon}{2} })
\otimes|s\rangle$, such that
\begin{equation}
\mathbb{P}(\epsilon, s,\ell)=
P(s, \ell) +  P(s, \ell + \frac{\epsilon}{2} ),
\qquad \epsilon= \pm 1,\qquad s=0,1,
\end{equation}
so that the projected coordinates onto this space
are expressed by
$\mathbb{P}(\epsilon, s,\ell)\, Z\,
  \mathbb{P}(\epsilon, s,\ell)= Z^{\epsilon, s,\ell}$,\,
  $Z={\mathcal X},\,{\mathcal Y}$.
Finally, the noncommuting coordinate algebra
associated with ${\mathcal E}_\ell$ can be written as,
\begin{eqnarray}
\textstyle\left[{\mathcal X}^{\epsilon, s,\ell},\,{\mathcal Y}^{\epsilon, s,\ell}
\right]=&\frac{i\hbar}{2m\Omega}&\textstyle
\left[
\sum_{k=0}^{\infty}
\{k+1\}_+ \left[|k+1,2\ell+k+{\frac{\epsilon + 3}{2}}, s\rangle
\langle k+1,2\ell+k+{\frac{\epsilon + 3}{2}}, s|\right. \right.
 \cr
&&\textstyle\left.    -  |k,2\ell+k+{\frac{\epsilon + 3}{2}}, s\rangle
\langle k,2\ell+k+{\frac{\epsilon + 3}{2}}, s| \right]\cr
&& \textstyle
+\sum_{k=0}^{\infty}
\{2\ell + k+{\frac{\epsilon + 3}{2}} \}_-
\left[|k,2\ell+k+{\frac{\epsilon + 1}{2}}, s\rangle
\langle k,2\ell+k+{\frac{\epsilon + 1}{2}}, s|\right.  \cr
&&\textstyle\left. -  |k,2\ell+k+{\frac{\epsilon + 3}{2}}, s\rangle
\langle k,2\ell+k+{\frac{\epsilon + 3}{2}}, s| \right]\cr
&& \textstyle
-\sum_{k=0}^{\infty}
\sqrt{\{k+1\}_+\{2\ell + k+{\frac{\epsilon + 5}{2}} \}_-}\cr
&&\textstyle
\left[|k,2\ell+k+{\frac{\epsilon + 3}{2}}, s\rangle
\langle k+1,2\ell+k+{\frac{\epsilon + 5}{2}}, s| \right. \cr
&&\textstyle  \left.  +  |k+1,2\ell+k+{\frac{\epsilon + 5}{2}}, s\rangle
\langle k,2\ell+k+{\frac{\epsilon + 3}{2}}, s| \right]\cr
&& \textstyle
+\sum_{k=0}^{\infty}
\sqrt{\{k+1\}_+\{2\ell + k+{\frac{\epsilon + 3}{2}} \}_-}
\cr
&& \textstyle
\left[|k+1,2\ell+k+{\frac{\epsilon + 3}{2}}, s\rangle
\langle k,2\ell+k+{\frac{\epsilon + 1}{2}}, s| \right.\cr
&&
 \textstyle\left.\left. +  |k,2\ell+k+{\frac{\epsilon + 1}{2}}, s\rangle
\langle k+1,2\ell+k+{\frac{\epsilon + 3}{2}}, s| \right]
\quad  \right].
\label{eq:su1}
\end{eqnarray}
The NC geometry thus appears in a different fashion
when the projection is done onto ${\mathcal E}_\ell$.
Alternatively, a nontrivial and different NC coordinate structure 
could also be investigated onto the associated
subspace of the $su(1,1)$ unitary representation space,
\begin{eqnarray} \textstyle
{\mathcal E}'_k= \left\{\;|k-1,\,2\ell+k\rangle, \;
\ell=-\frac{k}{2},-\frac{k}{2}+\frac{1}{2}, \ldots\;\right\},
\;\;
k\in\mathbb{N}/\{0\}.
\end{eqnarray}

\subsection{Reduced deformed model and supersymmetries}
\label{JBN.Subsect33}

Consider the Hamiltonian  ${\mathcal H}_{0}$ in (\ref{eq:hpqlan2})
under the set of conditions
$k=0=\omega$, $\vec{E}=\vec{0}$, $\Omega=\varpi=\omega_0/4=-K_0/\hbar$.
It can be rewritten as,
\begin{eqnarray}
{\mathcal H}_{0}'=
2\hbar\varpi (A^{\,\dag}_{-}A_{-} + [A_{-},A^{\, \dag}_{-}]\,b^{\, \dag}b).
\label{eq:hpri}
\end{eqnarray}
Therefore, defining
${\mathcal Q}_{-}=\kappa A^{\,\dag}_{-} b$ and
${\mathcal Q}_{-}^{\,\dag}=\kappa^* A_{-}b^\dag $,
with $\kappa=i\sqrt{2\hbar\varpi}$, one has
\begin{eqnarray}
\{ {\mathcal Q}_{-}, {\mathcal Q}_{-}^{\,\dag}\}= {\mathcal H}_{0}',\qquad
[{\mathcal Q}_{-}, {\mathcal H}_{0}'] =0,\qquad
[{\mathcal Q}_{-}^{\,\dag}, {\mathcal H}_{0}']= 0.
\label{eq:susy}
\end{eqnarray}
Thus, for this reduction
${\mathcal Q}_{-}$ and ${\mathcal Q}_{-}^{\,\dag}$
generate the well-known closed superalgebra
$sl(1|1)$\cite{coo}.
These are the $f$-deformed supercharges of the model
which generalise the ordinary charges
for the canonical Pauli--Landau operator.
Furthermore, according to the  projectors (\ref{eq:projectors})
and still in the notation of (\ref{eq:not}),
we have the following commutators
\begin{eqnarray}
\Big\{\overline{\mbox{\boldmath${\mathcal Q}$}}_{-},
\overline{\mbox{\boldmath${\mathcal Q}$}}_{-}^{\dag} \Big\}=
\overline{\mbox{\boldmath${\mathcal H}$}'}_{0},\qquad
\Big[\overline{\mbox{\boldmath${\mathcal Q}$}}_{-},
\overline{\mbox{\boldmath${\mathcal H}$}'}_{0} \Big]=0
=  \Big[\overline{\mbox{\boldmath${\mathcal Q}$}}_{-}^{\dag} ,
\overline{\mbox{\boldmath${\mathcal H}$}'}_{0} \Big],
\end{eqnarray}
showing that the superalgebra (\ref{eq:susy}) is conserved onto Landau
levels for the projected deformed charges.

\section{Concluding Remarks}
\label{JBN.Sec4}

Classes of $f$-deformations of Landau operator in a spherical harmonic well
within a superspace have been considered.
Even though the odd contribution of
the total Hamiltonian appears decoupled from the even sector
at the classical level, interesting features
have emerged through $f$-deformation
by coupling the bosonic and fermionic sectors.
It has been shown that the deformed problem is still exactly solvable
with more general underlying NC geometries
by considering various coordinate projections
onto two mode Fock and $su(1,1)$ infinite dimensional algebraic
representations. 
NC geometries of the deformed quantum Hall limit
have shown spin, potential and quantum algebra dependencies.
Finally, a $f$-modified version of a ${\mathcal N}=2$ superalgebra
has been discussed for a reduced model when the external electric field
and the harmonic oscillator well have been removed.

\section*{Acknowledgments}

This work is partially supported by the Abdus Salam International Centre for Theoretical Physics
(ICTP, Trieste, Italy) through the OEA-ICMPA-Prj-15 project. The ICMPA-UNESCO is in partnership with
the Daniel Iagoniltzer Foundation (DIF, France). J.B.G. is grateful to Prof. Hendrik Geyer 
for an invitation at the University of Stellenbosch (Republic of South Africa)
where this work has been finalised. He acknowledges the staff of Department of Physics
for their warm and welcoming  hospitality.

J.G. also acknowledges the ICTP Visiting Scholar Programme in support of
a Visiting Professorship at the ICMPA-UNESCO (Republic of Benin).
His work is also supported by the Institut Interuniversitaire des Sciences Nucl\'eaires,
and by the Belgian Federal Office for Scientific, Technical and Cultural Affairs through
the Interuniversity Attraction Poles (IAP) P6/11.

\end{document}